\documentclass[twocolumn,showpacs,preprintnumbers,amsmath,amssymb,twoside]{revtex4}

\usepackage{graphicx}
\usepackage{dcolumn}
\usepackage{bm}

\newcommand{\LMO}{LaMnO$_3$}

\newcommand{\LSMO}{La$_{\mbox{\scriptsize 7/8}}$Sr$_{\mbox{\scriptsize 1/8}}$MnO$_{\mbox{\scriptsize 3}}$}

\newcommand{\LSMOx}{La$_{\mbox{\scriptsize 1-x}}$Sr$_{\mbox{\scriptsize x}}$MnO$_{\mbox{\scriptsize 3}}$}
\newcommand{\mnk}{Mn K-edge}

\newcommand{\sipi}{$\sigma\pi$}
\newcommand{\sisi}{$\sigma\sigma$}
\newcommand{\tco}{T$_{\mbox{\scriptsize MI}}$}

\newcommand{\oo}{(1.5\,1.5\,3)}
\newcommand{\eg}{$e_g$}

\newcommand{\LCMOx}{La$_{1-x}$Ca$_{x}$MnO$_3$}
\newcommand{\PCMOx}{Pr$_{1-x}$Ca$_{x}$MnO$_3$}

\begin{document}


\title{Orbital Polaron Lattice Formation 
 in Lightly Doped \LSMOx}

\author{J. Geck$^{1,2}$}
\author{P. Wochner$^2$}
\author{S. Kiele$^{1,3}$}
\author{R. Klingeler$^{1}$}
\author{P. Reutler$^{1,4}$}
\author{A. Revcolevschi$^{4}$}
\author{B. B\"uchner$^{1}$}

\affiliation{$^1$Leibniz Institute for Solid State and Materials Reasearch  IFW Dresden, Helmholtzstr. 20, 01069 Dresden, Germany}


\affiliation{$^2$Max-Planck-Institut f\"ur Metallforschung, Heisenberg Str. 3, 70569 Stuttgart, Germany}

\affiliation{$^3$Hamburger Synchrotronstrahlungslabor HASYLAB at
 Deutsches Elektronen-Synchrotron DESY, Notkestr. 85, 22603 Hamburg,
 Germany}

\affiliation{$^4$Laboratoire de Physico-Chimie de l'Etat Solide, Universit\'e
  de Paris-Sud, 91405 Orsay Cedex, France}
%

\date{\today}

\begin{abstract}
By resonant x-ray scattering at the Mn K-edge on \LSMO, we show that an orbital polaron lattice (OPL) develops at the metal-insulator transition
of this compound. This orbital reordering explains consistently the unexpected coexistence of ferromagnetic and insulating properties at low
temperatures, the quadrupling of the lattice structure parallel to the MnO$_2$-planes, and the observed polarization and azimuthal dependencies.
The OPL is a clear manifestation of strong orbital-hole interactions, which play a crucial role for the colossal magnetoresistance effect and
the doped manganites in general.
\end{abstract}


\maketitle

During the last 20 years there have been tremendous efforts worldwide to unravel the physics behind the doping induced changes in magnetic
transition metal oxides. Besides the high-temperature superconducting copper oxides, where the interplay of spins and charges is relevant
\cite{OrensteinScience00}, the physics of doped manganites has become another main focus in this field. In these materials not only are the
spins and the charges important, but also the so-called orbital degree of freedom plays a central role for the physical properties
\cite{TokuraScience00}.
This is already documented by the insulating ground state of undoped \LMO, which does not only display AFM order but also antiferro-orbital
order, meaning that the spatial distribution of the $3d$-electrons alternates from one Mn-site to the next \cite{MurakamiPRL98}.  Doping of this
correlated magnet with holes results in a large number of intriguing physical phenomena. The most prominent is the colossal magneto-resistance
(CMR) effect, i.e. the enormous suppression of the electrical resistivity in applied magnetic fields.

On a qualitative level, the CMR can be understood in terms of the double exchange (DE) model. The DE mechanism refers to spin-charge
interactions only and implies that parallel spin alignment and charge itineracy are equivalent. Nonetheless, bare spin-charge scenarios fail to
describe the CMR effect on a quantitative level \cite{MillisPRL96}.
%
%
An even more clear example for the failure of bare spin-charge scenarios is the ferromagnetic insulating (FMI) phase of lightly doped \LSMOx\/
around the commensurate doping level x=1/8. The apparent contradiction between the FMI properties and the DE clearly indicates that the other
degrees of freedom referring to the lattice and the orbitals are indispensable to understand the physics of the manganite materials. In the case
of the FMI phase, this conjecture agrees with the observed complex structural modulations that point to a complex ordering of lattice and
orbital degrees of freedom \cite{TsudaJPSJ01,YamadaPRL96,InamiJJAP99,YamadaPRB00,NiemoellerEPJB99}.

Indeed, in a previous publication we could show that the development of the FMI phase of \LSMO\/ below the metal-insulator transition at \tco \/
is connected to an orbital reordering  \cite{geckPRB04}. But despite extensive theoretical and experimental efforts, there is no consensus about
the type of ordering, as manifested by the large variety of different models proposed in the literature
\cite{YamadaPRL96,InamiJJAP99,YamadaPRB00,EndohPRL99,MizokawaRPRB00,KorotinPRB00}.
In order to clarify the microscopic nature of the FMI phase and to identify the relevant interactions, we have performed resonant x-ray
scattering (RXS) experiments at the \mnk\/ on \LSMO.
Obviously, to be able to observe the complex orbital arrangement of only one electron per Mn$^{3+}$-site, ordinary diffraction is not
sufficient. By using in addition the spectroscopic signature of the Mn$^{3+}$ valence orbitals, appropriate positions in reciprocal space as
well as the dependence of the scattering intensity on the mutual orientation of x-ray polarization vector and orbital orientation, we achieve
enough sensitivity in this "spectro-diffraction", that we can unambiguously determine the long-range ordered orbital configuration.

The experimental data provides firm experimental evidence for the formation of an orbital polaron lattice (OPL) at low temperatures, which
unifies ferromagnetic and insulating properties in a natural way. To our knowledge, the OPL is the first microscopic model reproducing the
correct unit cell. Furthermore, the OPL in the FMI phase constitutes a clear manifestation of strong orbital-hole interactions, which play a
crucial role for the doping induced transition from an AFM insulating to a FM metallic state and are of crucial importance for the physics of
doped manganites in general.
%
%
%
%

%
%

In the course of the present RXS study, two \LSMO\/ single crystals with  carefully polished (001) and (112) surfaces have been investigated
($Pbnm$-setting).   The samples have been prepared from the same sample rod, obtained by the travelling floating zone method \cite{ReutlerCG02}.
Although there are six possible structural twin domains, whose [100], [010] and $\langle 112 \rangle$ directions can coincide, in our samples
only a subset of twins is realized, where [100] or [010] directions (type A twins) never overlap with $\langle 112 \rangle$ directions (type B
twins) \cite{geckPRB04}. Further experimental details about the RXS experiment, which has been performed at the wiggler beamline W1 at HASYLAB,
can be found in the literature \cite{geckPRB04}.

Upon cooling the samples from room temperature to 10 K, in addition to a symmetry change from orthorhombic to monoclinic and triclinic, several
types of superstructural modulations have been observed. The corresponding superstructure reflections can be grouped into I) so-called ATS
reflections, II) reflections with modulation wave vector (0,0,1/2), III) reflections with (1/4,-1/4,0), and IV) reflections with (1/2,1/2,0).
The ATS (Anisotropy of the Tensor of Susceptibility) reflections of I) are forbidden reflections due to glide plane and screw axis symmetries,
e.g. (300), (030), (003) or (104), which can be observed in resonance at the \mnk \/ 
\cite{DmitrienkoActaCryst83}. The reflections belonging to the groups II)-IV), appear only in the FMI phase below \tco$\simeq150$\,K. Group II)
reflections have been discussed in terms of charge order [8-10] and group IV) reflections in terms of an orbital rearrangement \cite{geckPRB04}.
\begin{figure}[t!]
\center{\includegraphics*[angle=0,width=0.7\columnwidth]{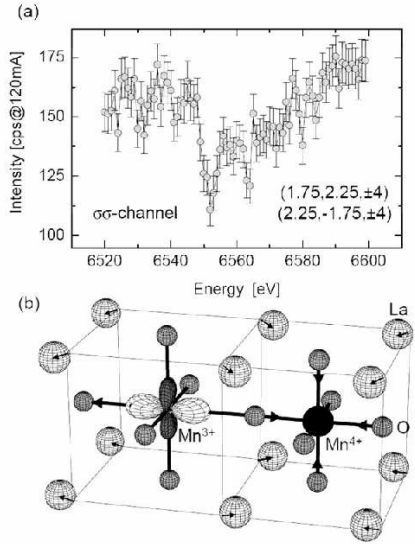}}
  \caption{(a): Energy dependence of the intensity at the nominal (1.75,2.25,4) position taken at T=50 K
  (without absorption corrections). (b) Illustration of a distortion that involves the heavy La-sites. The gray
  and white lobe represents the spatial $e_g$-electron distribution of a Mn$^{3+}$-site that corresponds
  to about 1/100 of the total charge per unit cell. The breathing
  distortion at the Mn$^{4+}$-site and shifts of the La-sites are indicated by arrows.} \label{fig1}
\end{figure}
\begin{figure}[t!]
\center{\includegraphics*[angle=-0,width=0.75\columnwidth]{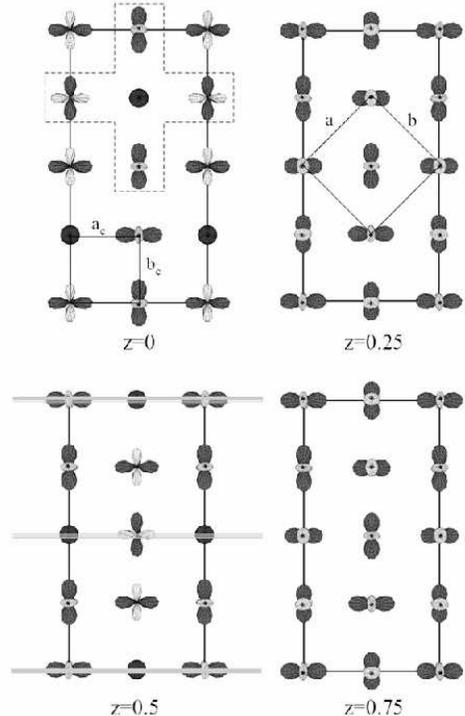}}
  \caption{Orbital polaron model  for the FMI-phase of
    \LSMO. A top view of the different $ab$-planes along the $c$-axis is
    shown. The orbitals have been calculated according to  the
    interaction $\mathcal H_{OP}$ with $\Delta/J=1$.
    The orthorhombic ($a$, $b$) as well as the pseudo-cubic ($a_c$, $b_c$)  lattice parameters are
    indicated. The hatched cross (z=0) and the gray bars (z=0.5) mark the orbital polaron and charge stripes in the $ab$-planes, respectively.}
\label{fig2}
\end{figure}
The studies on the (112)-oriented sample proved the occurrence of a superstructure reflection at the nominal (1.75,2.25,4) position in the FMI
phase (group III).
Since there is no superposition of type A and type B twins in our samples, the observation of the nominal (1.75,2.25,4) reflection unambiguously
shows that a quadrupling along the orthorhombic $[ 1 \bar 10 ]$ or $[ 110 ]$ exists below \tco. The energy dependence of the intensity at the
(1.75,2.25,4) position displays a dip-like feature around the Mn K-edge, as can be observed in Fig.\,\ref{fig1}\,(a). This implies a dominant
contribution to the intensity due to the heavy non-resonant scatterers, which participate in the corresponding superstructure modulation
(Fig.\,\ref{fig1}\,(b)). Note, that this type of energy dependence does not allow one to infer anything about possible charge and/or orbital
order, as has been done in the literature \cite{EndohPRL99}. Even in the naive case of ordered Mn$^{3+}$- and Mn$^{4+}$-ions, the corresponding
resonance can be easily masked by the signal of the heavier non-resonant sites.

To summarize so far, the present RXS measurements  reveal a superstructure modulation in the FMI phase, which involves a doubling along the
$c$-direction and a quadrupling parallel $[1 \bar 1 0]$ or $[110]$. Furthermore, the $(3/2,3/2,3)$ reflection (group IV) signals a doubling
along $[1 \bar 1 0]$ or $[110]$. The twinning prevents a unique identification, but because
the (1.75,2.25,4) reflection displays a dip at the \mnk, while a resonance is observed at the $(3/2,3/2,3)$ position (c.f. Ref.\,10), it is
tempting to assign them mutual orthogonal directions. This result is in agreement with a selected area electron diffraction (SAED) study which
probes  a single twin domain \cite{TsudaJPSJ01}. Both, the RXS and the SAED data imply a $2\, a_c \times 4\,b_c \times 4\, c_c$ unit cell in the
FMI phase, where the subscript $c$ indicates the pseudo-cubic lattice parameters of the high temperature orthorhombic phase (c.f.
Fig.\,\ref{fig2}). To the best of our knowledge, the quadrupling along the pseudo-cubic $b$-direction is not reproduced by any of the published
microscopic models for the FMI phase of \LSMO\/
\cite{YamadaPRL96,InamiJJAP99,YamadaPRB00,EndohPRL99,MizokawaRPRB00,KorotinPRB00,NiemoellerEPJB99}, which calls for the development of a model
based on the correct unit cell.
\begin{figure}[t!]
\center{\includegraphics*[angle=-0,width=0.72\columnwidth]{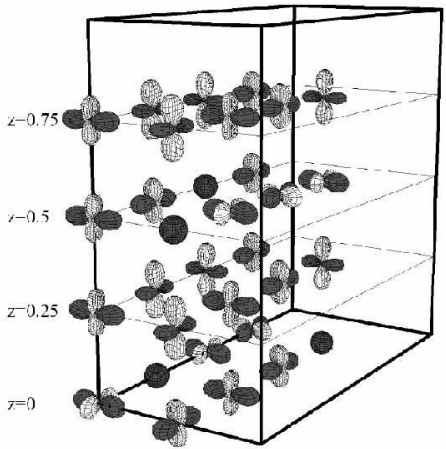}}
  \caption{Basis of the orbital polaron model for the FMI-phase of
    \LSMO which has been used for the calculation of the structure factor.
     The thick solid lines indicate the $2\,a_c \times 4\, b_c
    \times 4\, c_c$ supercell.}
\label{fig3}
\end{figure}
In a first approximation we will assume a strong localization of the doped holes, which is well verified by the pronounced insulating properties
of the FMI phase \cite{UhlenbruckPRL99}. Furthermore, we will omit the small triclinic distortion present below \tco\/ \cite{CoxPRB01}.

The superstructure modulation signaled by reflections of the type $(h,k,l+0.5)$, which are observed in neutron and x-ray scattering experiments,
as well as Hartree-Fock and LSDA+U calculations provide strong evidence for an alternation of hole-rich and hole-poor planes along the
orthorhombic $c$-direction below \tco \/  \cite{YamadaPRL96,YamadaPRB00,NiemoellerEPJB99,MizokawaRPRB00,KorotinPRB00}. The terms hole poor and
hole-rich indicate that there is not necessarily a perfect segregation into completely undoped and doped planes. Similarly, the terms hole-poor
and hole-rich Mn-sites will be used to stress that the valence difference $\delta q$ between Mn$^{3.5+\delta q}$ and Mn$^{3.5-\delta q}$ can be
less than $\delta q=0.5$. The theoretical model calculations also yield an antiferro-orbital ordering within the hole-poor planes
\cite{MizokawaRPRB00,KorotinPRB00}, indicating that the superexchange and/or electron-lattice interactions which stabilize this type of orbital
order \cite{FeinerPRB99} are still relevant in the FMI phase. In what follows, these interactions will be represented by the coupling constant
$J$. Hole induced interactions due to lattice distortions as well as Coulomb and double exchange interactions \cite{KilianPRB99,BalaPRB02b} will
be summarized by the coupling constant $\Delta$, hereafter. The interactions  $J$ and $\Delta$ both determine the orbital occupation at a given
hole-poor, i.e. Mn$^{3+}$-like site, next to a hole. In terms of a mean field approach, this can be modelled by Hamiltonians of the form
$\mathcal{H}_{OP}=-(J\,\sigma^x+\Delta\sigma^y$), which describe these interactions parallel to the $c$-direction ($\sigma^{x,y}$: Pauli
matrixes) \cite{KilianPRB99}.
Based on $\mathcal{H}_{OP}$, the alternation of hole-poor and hole-rich planes along the $c$-direction, and the $2\,a_c \times 4\, b_c
    \times 4\, c_c$ unit cell, we obtain a model for the FMI phase, which is depicted in Fig.\,\ref{fig2} and Fig.\,\ref{fig3} for $J/ \Delta=1$.
The interaction $J$ establishes an almost complete antiferro-orbital order within the hole-poor planes. However,  due to the additional
orbital-hole interaction $\Delta$ the antiferro-orbital ordering is modified. This is most evident in the hole-rich planes (z=0, 0.5) in
Fig.\,\ref{fig2}. $\Delta$ induces a polarization of the orbitals next to a hole and, consequently, the model constitutes an OPL.
\begin{figure}[t!]
\center{\includegraphics*[angle=-90,width=\columnwidth]{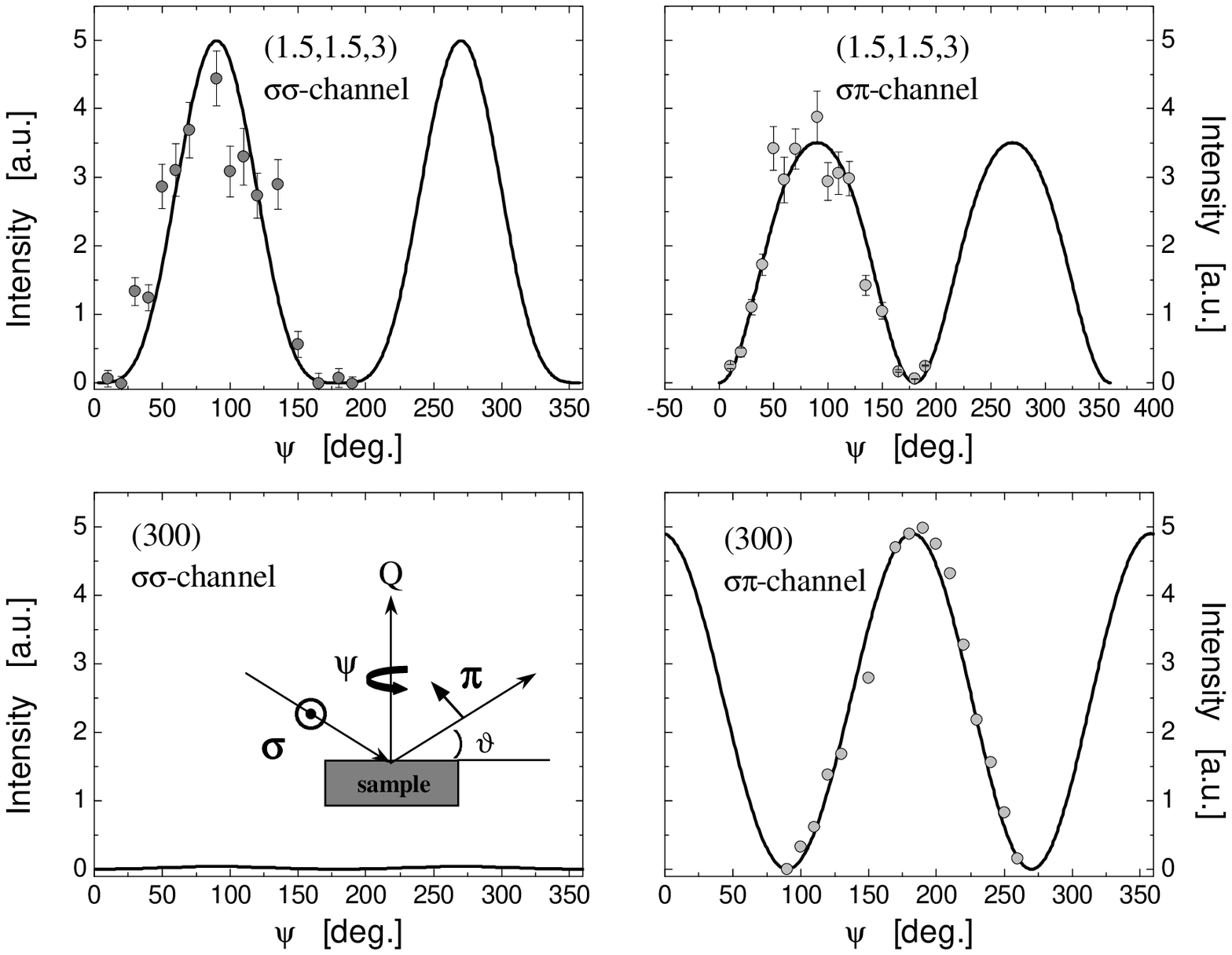}}
  \caption{Comparison between the measured data and a model calculation based on the orbital polaron lattice.
  The model calculation, which takes into account the twinning of the sample, reproduces the polarization dependencies observed at the nominal (300)
  and (1.5,1.5,3). A \sipi-scattering process is sketched in the lower left panel.}
\label{fig4}
\end{figure}

In order to check whether the ordering represented in Fig.\,\ref{fig2} 
is consistent with our experimental RXS results, we performed a model calculation based on the OPL. The sensitivity to orbital order of RXS at
the \mnk \/ stems from the fact that at this energy the scattering process involves Mn:4p-intermediate states. The energy splitting of the
Mn:4p-states reflects the orbital occupation at a given Mn-site $l$ and gives rise to a characteristic anisotropy of the corresponding form
factor $f_l$ with respect to the beam polarization \cite{MurakamiPRL98}. This polarization dependence can be described as
$f_l(\epsilon_i,\epsilon_f)=\epsilon_f^t.\hat f_l . \epsilon_i$, with the second rank tensor $\hat f_l$ and the polarization vectors
$\epsilon_i$ and $\epsilon_f$ of the incoming and outgoing beam, respectively. In the following $\sigma$- or $\pi$-polarization refer to
polarizations $\bot$ and parallel to the scattering plane, respectively (c.f. \cite{geckPRB04}).
%
%
For the hole poor sites we take the $\hat f_l$ to be diagonal with diagonal elements $f_l^{\alpha,\alpha}=f^{hp}(1+\delta_l^{\alpha})$
($\alpha=x,y,z$), where the $\delta_l^{x,y,z}$ are related to the splitting of the Mn:4p-states. As a result, the symmetry of the tensors $\hat
f_l$ corresponds to the symmetry of the orbitals in Fig.\,\ref{fig3}. The hole-rich sites are described by an isotropic $\hat f_l$; i.e
$f_l^{\alpha,\alpha}=f^{hr}$.
Using the set of basis sites shown in Fig.\,\ref{fig3} and the form factors described above, the structure factor matrix for a reflection $Q$
can be calculated: $\hat F(Q)=\sum_l \hat f_l \cdot e^{i d_l.Q}$, where $d_l$ describes the position of lattice site $l$. The resonantly
scattered intensity $I(Q)=|\epsilon_f^t. \hat F(Q) . \epsilon_i|^2$ is polarization dependent and displays a characteristic variation, when the
sample is rotated about $\psi$ around the scattering vector $Q$. This is the so-called azimuthal dependence. A straightforward calculation
yields the following expressions for the azimuthal dependencies of the (1.5 1.5 3) reflection:
\begin{eqnarray*}
I_{\sigma\pi} & = & A \sin^2 \vartheta \cos^2 \psi \sin^2 \psi + B \cos^2 \vartheta \sin^2 \psi\\
I_{\sigma\sigma} &= & A' \sin^4\psi + const.
\end{eqnarray*}

For the (300) reflection  
$I_{\sigma\pi}=A'' \cos^2 \vartheta \sin^2 \psi$  and $I_{\sigma\sigma}=0+\mathcal{O}(\delta^2)$
is obtained. In the above equations, $\vartheta$ is the scattering angle, the constants $A$, $A'$,$A''$ and $B$ are functions of the matrix
elements of the $\hat f_l$. In addition, the indices $\sigma$ and $\pi$ refer to the initial and final beam polarization (see Fig.\,\ref{fig4}).
The matrix elements have been used as parameters to fit the azimuthal dependencies of the (300) and the (1.5 1.5 3) simultaneously. As
demonstrated in Fig.\,\ref{fig4}, the OPL model nicely reproduces the polarization dependencies observed at the nominal (300) and \oo\/
position. In particular, the model is capable of reproducing the relative intensities observed at the (300) and the \oo\/ position in the
\sisi-\/ and the \sipi-channel, although it has to be noted that the fitted components of the $\hat f_i$, could not be determined unambiguously.
But we emphasize that the simultaneous description of the data in Fig.\,\ref{fig4} provides very strict constraints for the orbital polaron
order. For example, an OPL based on the charge ordering pattern proposed in Ref.\,\cite{YamadaPRL96} always leads to
$I_{\sigma\sigma}=\mathcal{O}(\delta)$ for the (300) reflection; i.e. there is always a significant intensity in the \sisi-channel.
The ordering shown in Fig.\,\ref{fig2} is similar to one of the ground states obtained by Hartree-Fock calculations for \LSMO\/
\cite{MizokawaRPRB00}, supporting the scenario developed above. Note, that orbital polarons are ferromagnetic objects which lead to a strong
reduction of the charge carrier bandwidth \cite{KilianPRB99}.
A closer look at the OPL shows that there is always a non-vanishing overlap between an empty and an occupied \eg-state; i.e. the resulting
exchange interactions are all ferromagnetic. As a result, the OPL unifies ferromagnetic and insulating properties. Interestingly, according to
this model half doped charge stripes exist below \tco. Since these stripes run along the $a_c$-direction, which is doubled below \tco, one may
speculate whether a mechanism similar to a Peierls effect is relevant for the FMI phase.

It is important to note, that there is clear experimental evidence that local charge hopping (DE) processes contribute significantly to the
stabilization of the FMI phase \cite{geckNJP04,KlingelerPRB01}. This is in perfect agreement with the OPL model, because orbital
polarons are stabilized by local DE processes. 
Furthermore, the phase diagram of \LSMOx\/ as a function of $x$ can be reproduced by a model where the OPL formation competes with the
AF-orbital order and with orbital fluctuations \cite{KilianPRB99}. This competition together with the importance of local DE processes also
explains the stability of the OPL in other compounds.
For instance, in lightly doped \LCMOx\/ and \PCMOx \/ the DE interaction is reduced and, correspondingly, the AF-orbital ordering is observed
instead of the OPL \cite{ZimmermannPRB01,JirakJMMM85}. Nonetheless, the tendency to form orbital polarons is also expected to be present in
these systems.
We emphasize that the OPL has been derived based on the interactions described by $\mathcal{H}_{OP}$, which means that the OPL is a direct
manifestation of strong orbital-hole interactions in doped manganites.
Furthermore, it can be concluded that these interactions are strongly spin-dependent, since the DE mechanism is found to contribute
significantly to the stabilization of FMI phase \cite{KlingelerPRB01}. According to the above discussion, the spin-dependent orbital-hole
interaction is a general feature of doped manganite compounds and it is clear that such an interaction is of great relevance for the CMR effect.

We are very grateful to H. Dosch for his support. Furthermore, the authors would like to thank M. v. Zimmermann for fruitful discussions.

\end{document}